\documentclass[11pt,twocolumn,twoside]{article}

\usepackage{times}            
\usepackage{authblk}          
\usepackage{graphicx}         
\usepackage{amsmath,amssymb}  
\usepackage{xcolor}           
\usepackage{hyperref}         

\newcommand{\orgdiv}[1]{#1}
\newcommand{\orgname}[1]{#1}
\newcommand{\orgaddress}[1]{#1}

\newcommand{\postcode}[1]{#1}
\newcommand{\state}[1]{#1}
\newcommand{\country}[1]{#1}

\usepackage{fancyhdr}
\title{\textbf{inMOTIFin: a lightweight end-to-end simulation software for regulatory sequences}}
\author[1]{Katalin Ferenc}
\author[1,2]{Lorenzo Martini}
\author[1]{Ieva Rauluseviciute}
\author[3]{Geir Kjetil Sandve}
\author[1,4,5,$\ast$]{Anthony Mathelier}

\affil[1]{\small{\orgdiv{Norwegian Centre for Molecular Biosciences and Medicine (NCMBM), Nordic EMBL Partnership}, \orgname{University of Oslo}, \orgaddress{\postcode{0318}, \state{Oslo}, \country{Norway}}}}
\affil[2]{\small{\orgdiv{Department of Control and Computer Engineering (DAUIN)}, \orgname{Politecnico di Torino}, \orgaddress{\postcode{10129}, \state{Turin}, \country{Italy}}}}
\affil[3]{\small{\orgdiv{Department of Informatics}, \orgname{University of Oslo}, \orgaddress{\state{Oslo}, \country{Norway}}}}
\affil[4]{\small{\orgdiv{Department of Medical Genetics, Institute of Clinical Medicine}, \orgname{Oslo University Hospital and University of Oslo}, \orgaddress{\state{Oslo}, \country{Norway}}}}
\affil[5]{\small{\orgdiv{Bioinformatics in Life Science (BiLS) initiative, Department of Pharmacy}, \orgname{University of Oslo}, \orgaddress{\state{Oslo}, \country{Norway}}}}

\date{}

\begin{document}
\twocolumn[
\maketitle

\begin{abstract}
The accurate development, assessment, interpretation, and benchmarking of bioinformatics frameworks for analyzing transcriptional regulatory grammars rely on controlled simulations to validate the underlying methods. However, existing simulators often lack end-to-end flexibility or ease of integration, which limits their practical use. We present inMOTIFin, a lightweight, modular, and user-friendly Python-based software that addresses these gaps by providing versatile and efficient simulation and modification of DNA regulatory sequences. inMOTIFin enables users to simulate or modify regulatory sequences efficiently for the customizable generation of motifs and insertion of motif instances with precise control over their positions, co-occurrences, and spacing, as well as direct modification of real sequences, facilitating a comprehensive evaluation of motif-based methods and interpretation tools. We demonstrate inMOTIFin applications for the assessment of \textit{de novo} motif discovery prediction, the analysis of transcription factor cooperativity, and the support of explainability analyses for deep learning models. inMOTIFin ensures robust and reproducible analyses for studying transcriptional regulatory grammars.\\

inMOTIFin is available at PyPI \url{https://pypi.org/project/inMOTIFin/} and Docker Hub \url{https://hub.docker.com/r/cbgr/inmotifin}. Detailed documentation is available at \url{https://inmotifin.readthedocs.io/en/latest/}. The code for use case analyses is available at \url{https://bitbucket.org/CBGR/inmotifin_evaluation/src/main/}.\\
\end{abstract}

\noindent \textit{keywords}: gene regulation | simulation | transcription factors | regulatory grammar

\vspace{.5em}
\noindent \textit{corresponding author:} anthony.mathelier@ncmbm.uio.no
]

\section*{Introduction}\label{sec1}

When developing algorithms or computational models, it is necessary to evaluate their performance in a well-controlled setting \cite{sandve_access_2022}.
Although real biological sequences serve as the ultimate test of performance and application, they are often limited in numbers, and one lacks the underlying ground truth they contain.
As such, they might not correspond to a gold standard and are often inadequate for thoroughly evaluating a model's assumptions and behavior in various scenarios, including edge cases.

In the context of regulatory genomics, the goal is to decipher the cis-regulatory code, which governs transcription control through the specific binding of transcription factors (TFs) at binding sites \cite{zeitlinger_perspective_2025}.
Specifically, TFs recognize DNA sequence patterns and act cooperatively at cis-regulatory regions (such as promoters, enhancers, and silencers).
Several computational methods have been developed over the years to identify motifs bound by TFs, which are embedded in the DNA sequences of cis-regulatory regions, as well as how their composition regulates transcription.
For instance, common tasks consist in discovering TF binding patterns {\em de novo} within experimentally identified sequences \cite{van_helden_extracting_1998}, elucidating the cooperativity of binding between pairs of TFs \cite{rauluseviciute_identification_2024}, or, more recently, to interpret deep learning models that predict the activity or function of DNA sequences \cite{sasse_unlocking_2024, zeitlinger_perspective_2025}. 
As the underlying ground truth is usually unknown, simulated motifs and DNA sequences are commonly used to assess computational tools during their development and subsequent benchmarking \cite{tompa_assessing_2005, gupta_quantifying_2007, simcha_limits_2012, shrikumar_learning_2019, rauluseviciute_identification_2024}.
Real biological sequences are often modified to support model evaluation, especially in the case of explainability tools for deep learning models \cite{gupta_quantifying_2007, pmlr-v165-prakash22a}.

However, to our knowledge, no single package provides a self-contained, easy-to-use, end-to-end simulation of regulatory sequences by implanting or modifying motif instances (or sites) and grammar in user-provided or simulated background sequences.
To address this gap, we developed inMOTIFin, a flexible Python package and command-line tool designed to simulate DNA motifs and background sequences, and to insert motif instances following user-defined grammars. Distinctively, and unlike existing simulators (Supplementary Table~1), inMOTIFin integrates flexible and customizable motif grammars, direct sequence modification capabilities, and a modular architecture, thereby facilitating comprehensive benchmarking and systematic evaluation of regulatory genomics software.
Furthermore, inMOTIFin allows for the creation of DNA sequences that contain instances of motifs distributed in groups to simulate cis-regulatory grammars through cooperative binding of TFs. Finally, the package supports the modification of real biological sequences, such as modifying single nucleotides, masking out existing motif sites, or shifting the position of a motif site on a sequence-by-sequence basis.
Within inMOTIFin, these simulation tasks are easily parameterized for various use cases.
We demonstrate how to utilize inMOTIFin through three use cases: simulating motifs of different lengths and information content, inserting dimers into user-provided or random sequences, and masking existing motif instances in sequences to support the explainability of deep learning models.

\section*{Features and implementation}\label{sec11}

\subsection*{Main feature: simulation of sequences with a grammar of TF binding motif instances}

The primary feature of inMOTIFin is the simulation of DNA sequences with motif instances inserted following a grammar of co-occurring motifs.
Users can define the motif grammar to specify the spacing between co-occurring motifs (see Section Additional features).
Alternatively, for a more flexible grammar, motifs are organized into groups before being inserted into sequences (see Supplementary Section~3).

For a simulation task, the user prepares a single configuration file that contains a set of parameters controlling the simulation and sampling of background sequences and motifs, as well as the number, orientation, and positional distribution of motif instances in the final sequences.
A comprehensive documentation of the parameters is available at \url{https://inmotifin.readthedocs.io/en/latest/usage/command\_line\_options.html}.
inMOTIFin starts by preparing data collections and distributions to sample from.
First, a pool of background sequences and a pool of motifs are created.
Motifs are then organized into groups.
Next, the distributions for sampling are set up. Specifically, the probability of selecting a motif is defined as a set of multinomial distributions, which are specified at both the group level (including the occurrence and co-occurrence of groups) and the motif level within groups.
The orientation of the inserted motif instance follows a binomial distribution with a user-provided probability of success.
The positions of the instances can be either fixed in the middle or sampled from a uniform or a Gaussian distribution.
The number of instances per sequence may be a specific user-provided value or drawn from a Poisson distribution.
Note that inMOTIFin supports direct user input for each step, where appropriate, allowing, for example, the modification of real biological sequences with known TF binding motif instances.
Following this initial preparation phase, sampling takes place.

During sampling, each component of the simulation is sampled from its respective distribution.
To achieve this, inMOTIFin is built on top of DagSim, a simulation framework for causal models \cite{Hajj_dagsim_2023}.
The complete model is represented as a directed acyclic graph with nodes defining random processes or user input (Supplementary Figure~1).
The components of the simulation are the selection of groups of expected co-occurring motifs, the choice of motifs, the choice of background sequences, the sampling of motif instances, the sampling of the number of groups and motif instances per sequence, and the sampling of the positions and orientations of the motif instances.
The simulated values are passed to their downstream counterparts in each round, resulting in one output sequence at a time.
The total number of rounds corresponds to the number of expected sequences.

Through the Python interface, users have access to more features, most notably better control over the simulation.
This includes setting the exact position for each motif instance in each sequence, as well as the option to mask out existing motif sites in DNA sequences with any nucleotide (including Ns).
For examples, visit \url{https://inmotifin.readthedocs.io/en/latest/usage/python_module.html}.

\begin{figure}
    \centering
    \includegraphics[width=1\linewidth]{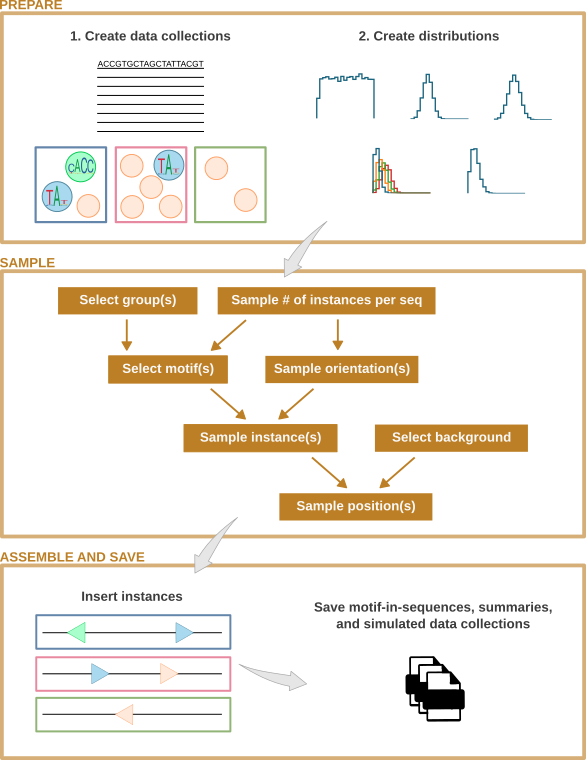}
    \caption{\textbf{inMOTIFin simulation framework:} In preparation, background and motif pools are created, and motifs are assigned to groups. Distributions are set up for sampling of various entities. In the sampling phase, backgrounds, motif groups, and subsequently motifs are sampled. Two motifs are more likely to co-occur in a sequence if assigned to the same group. When motifs from multiple groups can be inserted into the same sequence, the group-group co-occurrence probabilities are conditioned on the first selected group. Instances are sampled from motifs and inserted into the background sequences in the selected orientations and positions. All newly created data is saved into files, along with additional files to facilitate downstream analyses.}
    \label{fig:inm_summary}
\end{figure}

The package outputs a FASTA file with all the sequences with motif instances inserted. Moreover, complementary output files are provided to facilitate interoperability with downstream tools:
\begin{enumerate}
    \item a BED file providing the positions of the inserted motif instances in the sequences;
    \item a table in CSV format that includes all the details about the simulated sequences, motifs, instances, orientations, and positions;
    \item a JSON file with counts for all sampled entities: backgrounds, groups, motifs, orientations, sequence of motif instances, number of instances per sequence, positions, and motif lengths;
    \item a FASTA file providing all simulated background sequences (if applicable);
    \item a MEME file with the simulated motifs (if applicable);
    \item four TSV files detailing the probability of motif-motif co-occurrences produced (if applicable).
\end{enumerate}

\subsection*{Additional features}

In addition to the end-to-end simulation functionality, inMOTIFin enables the specific simulation of random sequences and TF binding motifs.
To generate random background sequences, inMOTIFin requires the user to provide the sequence length, the number of sequences, the alphabet, and the frequency of each letter in the alphabet.
When generating TF binding motifs, the parameters are the minimum and maximum length of the motifs (the motif lengths are sampled uniformly between the minimum and maximum lengths), the number of motifs, the alphabet, and the $\alpha$ values for each letter in the Dirichlet distribution. Indeed, inMOTIFin relies on the Dirichlet distribution to control the average information content per position in the motif (see \url{https://inmotifin.readthedocs.io/en/latest/usage/motif_simulation.html} for a detailed explanation and examples).
Moreover, inMOTIFin enables the simulation of motifs corresponding to multimers (e.g., fixed spacing between a pair of motifs).
This can be achieved by providing a file with a set of motifs and a file with the multimerization rules (see \url{https://inmotifin.readthedocs.io/en/latest/usage/command_line_options.html#multimerisation-of-motifs}). Note that the spacings may be negative, allowing for compressed motifs or "noised" motifs when a random signal is combined with a known motif.

\subsection*{Implementation}

InMOTIFin is implemented as both a command-line tool and a Python package.
The Python package provides access to the individual classes and functions that comprise the simulation framework. The time performance of the simulation modules is O(n) (see Supplementary Figures~7-9).
The implementation is modular and well-documented, allowing users to utilize polymorphism by defining their classes and functions as replacements for any component (see the complete documentation at \url{https://inmotifin.readthedocs.io/en/latest/usage/python_module_usage.html}).

\section*{Use cases}\label{sec2}

To illustrate the capability of inMOTIFin to generate DNA sequences following a defined grammar of co-occurring motifs, we produced 50,000 random sequences with specific cis-regulatory grammar properties.
We aimed to generate sequences that include instances of eight distinct TF binding motifs grouped into three categories for co-occurrence: motifs 0–2 in group 0, motifs 2–5 in group 1, and motifs 1 and 6-7 in group 2 (see Supplementary Section~3 for detailed probabilities).
A pairwise intersection analysis of the sequences containing instances of the eight motifs confirmed that the generated sequences correctly incorporated motif instances according to the specified soft syntax regulatory grammar (Supplementary Figure~2).

To further demonstrate possible applications of inMOTIFin, we present three illustrative use cases. 
These examples highlight some of the potential uses of inMOTIFin rather than providing a comprehensive evaluation or benchmarking against downstream tools used on the simulated DNA sequences.

\subsection*{Use case 1: de novo motif discovery}

This use case involves inserting instances of simulated TF binding motifs into DNA sequences for application with \textit{de novo} motif discovery tools.
In Supplementary Section~4.1, we demonstrate how a {\em de novo} motif discovery tool can be evaluated for its capacity to uncover enriched motifs of varying lengths and information content (IC) generated by inMOTIFin and inserted in random background sequences.
By varying the length and IC of simulated motifs, one can determine the optimal detection boundaries of a given tool.
In this experiment, we specifically used RSAT's \textit{de novo} discovery algorithms \cite{santana-garcia_rsat_2022} and compared the extracted motifs with the inserted ground truth motifs using Tomtom \cite{gupta_quantifying_2007}.
As expected, long and high IC motifs are consistently identified, while shorter or lower IC motifs are more frequently overlooked by the tool (Supplementary Figure~3).
We demonstrate that the lowest threshold for discovery is a length of 6 (corresponding to the user-defined setting of RSAT) and an average IC of 1.4 per position.
Shorter motifs can be discovered if the IC is high, and conversely, low IC motifs can be discovered if they are long. This demonstrates how inMOTIFin reliably supports the assessment of \textit{de novo} motif discovery tools, identifying potential limits of tool sensitivity.

\subsection*{Use case 2: generating sequences with co-occurring motifs reflecting pairs of cooperative TFs}

To enable complex transcriptional regulation mechanisms, some TFs bind as complexes to cooperatively control transcription \cite{slattery2014absence}.
The inMOTIFin package supports the simulation of multimer motifs and the insertion of groups of motifs within DNA sequences.
In this use case, we demonstrate that inMOTIFin can simulate a set of dimer motifs with a user-selected composition of the two individual TF binding motifs that compose the dimers, corresponding to a hard syntax regulatory grammar (see Supplementary Section~4.2).
We simulated a set of sequence datasets with various dimers inserted in a defined proportion of the complete set of sequences.
Anchoring our sequences on the instances of a defined inserted motif, we applied SpaMo \cite{whitington_inferring_2011} to identify the secondary motifs that compose the dimers.
With sequence simulation, we can test how well these different motif combinations can be found (Supplementary Figure~4).
Overall, we observed that the motif identity within the pairs does not have an effect, as all motifs are identified.
In contrast, the accuracy of recovery for the specific sites of insertion increases when more motif combinations are present in a single dataset  (Supplementary Figure~4). Thus, inMOTIFin robustly facilitates the evaluation of cooperative motifs and motif instance detection under various regulatory complexities.

\subsection*{Use case 3: explainability of deep learning models}
\label{sec:usecase3}
The development and use of deep learning models are rapidly growing in genomic research, with an increasing number of frameworks being developed to perform various analyses \cite{Alharbi2022, zeitlinger_perspective_2025}.
Among them, sequence-based models are particularly relevant, where inputs are genomic sequences \cite{Barbadilla2025} and models are trained to infer specific functions, particularly pertinent to unravel transcriptional grammar \cite{zeitlinger_perspective_2025}.
In this context, inMOTIFin has the potential to support motif-centered research through easy integration with sequence-based deep learning architectures, thereby enhancing model explainability \cite{Bhagya2023}.

A prime example is the input perturbation approaches \cite{IVANOVS2021}, which involve selectively modifying an input DNA sequence and measuring the resulting change in the output function.
In the case of sequence-based models, inMOTIFin assists in the automated creation of both background sequences (respecting specific conditions, such as \%GC content) and their perturbation.
Specifically, inMOTIFin allows for the controlled insertion of motif instances to assess changes in the model output.
Alternatively, the masking functionality can replace known motif instances in real sequences with generic backgrounds, or even specifically mutate single nucleotides.
Supplementary Section~4.3 shows the application of a simple sequence-based model for detecting the GATA and TAL motifs.
The results demonstrate how the insertion of instances in background sequences alters the model output, highlighting the selective sensitivity of the model to different motifs (Supplementary Figures~5-6). Hence, inMOTIFin enables advanced interpretability efforts to elucidate model interpretation and sensitivity.

\section*{Conclusion}\label{sec12}

Simulation is an important, although time-consuming, part of computational software development.
The lightweight, robust, easily integrable, and automated software tool we present here will aid tool evaluation and benchmarking.
InMOTIFin is an open-source software built in a modular way to support extensibility, with a focus on the creativity of future users.
Beyond the use cases outlined here, there are many other possible applications, including model selection through the simulation of novel types of data based on explicitly stated assumptions; replacement of PWM-based simulation with CWM or seqlet-based \cite{seqlet_2024} simulation; or simulation of sequences using an extended alphabet \cite{viner_modeling_2024}.
With inMOTIFin, we aim to pave the way for a more thorough evaluation of models used in the biological context, where ground truth is often lacking and confounding factors can mask signals.

\section*{Competing interests}

No competing interest is declared.

\section*{Author contributions}

We follow here the Contributor Roles Taxonomy (CRediT) \cite{brand_beyond_2015}. KF: conceptualization, methodology, software, visualization, project administration, writing – original draft; LM: methodology, software, visualization, writing – original draft; IR: methodology, visualization, writing – original draft; GKS: conceptualization, supervision, writing - review \& editing; AM: conceptualization, writing - review \& editing, supervision, funding acquisition.

\section*{Acknowledgments}

We thank the members of the Software Quality seminars for providing feedback on the project, Stefano Di Carlo for reviewing the manuscript and offering valuable feedback, and the members of the Kuijjer and Mathelier groups for insightful discussions.

This work is supported by funding from the Research Council of Norway [187615], the Helse SørØst, and the University of Oslo through the Norwegian Centre for Molecular Biosciences and Medicine (NCMBM; formerly NCMM) [to Mathelier group], and the Norwegian Cancer Society [215027, 272930 to Mathelier group].

\bibliographystyle{unsrt}
\bibliography{inmotifin}

\end{document}


\maketitle

\section{Sampling procedure}

\begin{figure}[tbh]
    \centering
    \includegraphics[width=1\linewidth]{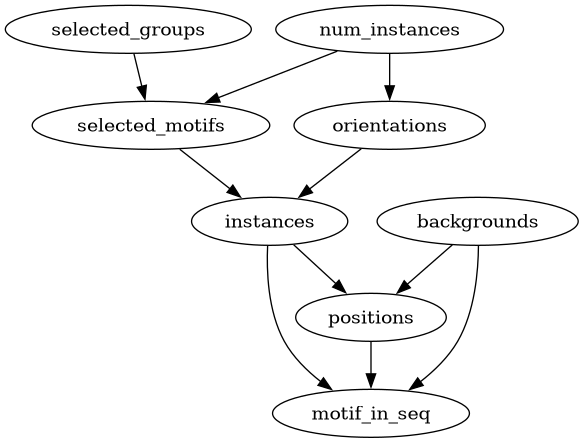}
    \caption{\textbf{Overview of the directed acyclic graph supporting the simulation framework in inMOTIFin.}}
    \label{fig:inm_dag}
\end{figure}

The complete direct acyclic graph supporting the simulation framework implemented in DAGSim \cite{Hajj_dagsim_2023} for inMOTIFin is illustrated in Supplementary Figure~\ref{fig:inm_dag}. The number of motif instances (\textit{num\_instances}) to be inserted into the background sequences can be set as a single integer or can be sampled from a Poisson distribution.
The motif position probability matrices are selected with replacement from a list of motifs (\textit{selected\_motifs}), which can be read from a file, accessed directly from the JASPAR database \cite{rauluseviciute_jaspar_2024} using the pyJaspar package \cite{khan_pyjaspar_2021}, or sampled from a user-defined Dirichlet distribution.
The orientations (\textit{orientation}) are sampled from a binomial distribution.
Instances (\textit{instances}) are sampled from the motif on the fly, given the alphabet of the sequences.
The background sequences (\textit{background\_id}) are sampled, either from a list of sequences provided by the user or generated by sampling from a distribution where the user provides the probability of each letter.
The positions (\textit{positions}) where the motif instances are inserted within the background sequences are centered or sampled from a uniform or a Gaussian distribution.
The final sequence (\textit{motif\_in\_seq}) is assembled using the selected background sequence, the selected motif instances, and their selected positions and orientations.
Motifs are organized into groups, allowing groups of motifs to be preferentially selected together (\textit{selected\_groups}) based on predefined or simulated group frequencies and co-occurrence probabilities.

\section{Comparison with other tools}

Several existing packages support the simulation of motifs, motif instances, background sequences, or the insertion of motif instances into background sequences; their key features are summarized in Supplementary Table~\ref{tab:comparison}. Tools such as RSAT, universalmotif, and the JASPAR web interface implement random motif generation in various ways. Both RSAT and universalmotif allow sampling from user-provided motif instances, while RSAT, universalmotif, and BPNet can generate background sequences. Additionally, RSAT, BPNet, and the discontinued rMotifGen tool support the insertion of motif instances into background sequences. inMOTIFin integrates all these functionalities into a single lightweight and automated solution, providing more flexibility and ease-of-use compared to existing tools.

\begin{table}
\begin{tabular}{ |P{2.5cm}||P{4cm}|P{4cm}|P{4cm}| }
 \hline
   & motif and motif instance simulation & background simulation & insertion of motif instances into background sequences \\
  \hline
  JASPAR \cite{rauluseviciute_jaspar_2024} & \textit{permute} columns of motifs or \textit{randomize}, which generates motifs similar to the ones in a previous version of the database & NA & NA \\
 \hline
 RSAT \cite{santana-garcia_rsat_2022} & \textit{random-motif} randomly selects one nucleotide from ACGT for each position, which gets a user-defined value, and the rest are of equal value; \textit{random-sites} samples motif instances from motifs & \textit{random-seq} creates sequences from length, number, and frequencies input; supports Markov-chains from specified organism and various nucleotide sizes & \textit{implant-sites} inserts provided instances into provided backgrounds \\
 \hline
  universalmotif \cite{tremblay_universalmotif_2024} & \textit{create\_motif} from a non-informative Dirichlet distribution and user-provided background frequencies of nucleotides; \textit{shuffle\_motifs} shuffles motifs by columns; \textit{sample\_sites} samples motif instances from motifs & \textit{create\_sequences} creates sequences from length, number, and frequencies input & NA \\
 \hline
 BPNet \cite{avsec_base-resolution_2021} & NA & \textit{random\_seq} generates a sequence with ACGT bases selected with equal probability for each position & \textit{insert\_motif} inserts a single instance of a motif centering at user provided position; \textit{generate\_seq} inserts two motif instances, one at the center and one at user provided distances from the centered one \\
 \hline
 rMotifGen* \cite{rouchka_rmotifgen_2007} & NA & NA & user can provide nucleotide frequencies and motifs, the output is a random sequence with motif instances and their locations \\
 \hline
 inMOTIFin & creates random motifs based on length range and alpha values for Dirichlet prior from which motifs are generated & creates sequences from provided length, number, and frequencies & single parametrized framework to connect all tasks and allow control for selecting number of motifs and instances, co-occurrence frequency of motifs, positions and orientations to build the final sequence \\
 \hline
\end{tabular}
\caption{\textbf{Comparison of features with other simulation tools}. *Assessed from the manuscript, code and GUI not available anymore.}
\label{tab:comparison}
\end{table}

\clearpage

\section{Grammar of motif co-occurrence simulation}
\label{Suppl:Cooccurrence}

The inMOTIFin tool enables motifs to be organized into groups, each with its own defined probability of being selected. Within each group, motifs are selected according to specified probabilities. Additionally, users can define probabilities for the co-occurrence of different motif groups. Here, we illustrate how this setup allows for the precise control and insertion of co-occurring motifs into sequences.

Four files are provided for this simulation and contain:
\begin{itemize}
    \item the probabilities of selecting each group for insertion (Table~\ref{tab:group_freq}).
    \item the conditional probabilities for selecting additional groups after an initial group is chosen, applicable when sequences contain multiple groups (Table~\ref{tab:group_pairs}). For this specific simulation, each sequence contains only one group, making this file ignored.
    \item the assignment of individual motifs to their respective groups (Table~\ref{tab:motif_in_group}).
    \item the probabilities of selecting specific motifs within each group (Table~\ref{tab:motif_prob}), which determines the co-occurrence frequency of motif pairs.
\end{itemize}

Further details can be found at \url{https://inmotifin.readthedocs.io/en/latest/usage/command_line_options.html}.

\begin{figure}[tbh]
    \centering
    \includegraphics[width=1\linewidth]{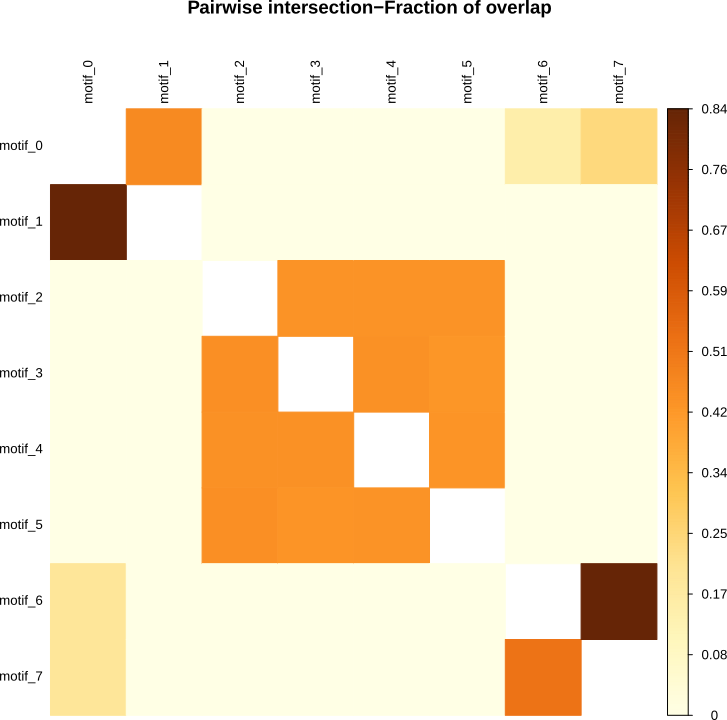}
    \caption{\textbf{Pairwise occurrences of inserted motifs within the simulated sequences}. The heatmap provides the frequencies of co-occurrence of the motif instances within the simulated sequences, confirming the regulatory grammar of the groups provided as input to inMOTIFin.}
    \label{fig:pairwise_fractions}
\end{figure}

The assignment of the underlying groups making up the cis-regulatory grammar was assessed using Intervene \cite{khan_intervene_2017} to confirm the co-localization of motif instances between within and between groups (Supplementary Figure~\ref{fig:pairwise_fractions}).

\begin{table}[b]
\begin{tabular}{ |P{5cm}|P{5cm}| }
 \hline
 Group Name & Probability \\
 \hline
   group\_0 & 0.2 \\
  \hline
  group\_1 & 0.3 \\
 \hline
 group\_2 & 0.5 \\
 \hline
\end{tabular}
\caption{\textbf{Motif group probability values}. This is similar to the example file \textit{group\_freq\_file.tsv} available in the documentation.}
\label{tab:group_freq}
\end{table}

\begin{table}[b]
\begin{tabular}{ |P{3cm}|P{3cm}|P{3cm}|P{3cm}| }
 \hline
   & group\_0 & group\_1 & group\_2 \\
  \hline
  group\_0 & 1 & 0 & 0 \\
 \hline
 group\_1 & 0 & 1 & 0  \\
 \hline
 group\_1 & 0 & 0 & 1  \\
 \hline
\end{tabular}
\caption{\textbf{Group pair probabilities}. This is similar to the example file \textit{group\_group\_file.tsv} available in the documentation.}
\label{tab:group_pairs}
\end{table}

\begin{table}[b]
\begin{tabular}{ |P{5cm}|P{5cm}| }
 \hline
 Group Name & Motifs \\
 \hline
   group\_0 & motif\_0, motif\_1 \\
  \hline
  group\_1 & motif\_2, motif\_3, motif\_4, motif\_5 \\
 \hline
 group\_2 & motif\_6, motif\_7, motif\_1 \\
 \hline
\end{tabular}
\caption{\textbf{Motif assignments to the groups}. This is similar to the example file \textit{group\_motif\_assignment\_file.tsv} available in the documentation.}
\label{tab:motif_in_group}
\end{table}

\begin{table}
\begin{tabular}{ |P{3cm}|P{3cm}|P{3cm}|P{3cm}| }
 \hline
 &  group\_0 &  group\_1 &  group\_2 \\
 \hline
   motif\_0 & 0.6 & 0 & 0.1 \\
  \hline
  motif\_1 & 0.4 & 0 & 0 \\
 \hline
  motif\_2 & 0 & 0.25 & 0 \\
 \hline
 motif\_3 & 0 & 0.25 & 0  \\
 \hline
 motif\_4 & 0 & 0.25 & 0  \\
 \hline
 motif\_5 & 0 & 0.25 & 0  \\
 \hline
 motif\_6 & 0 & 0 & 0.3 \\
 \hline
 motif\_7 & 0 & 0 & 0.6 \\
 \hline
\end{tabular}
\caption{\textbf{Motif probabilities within the groups}. This is similar to the example files \textit{called motif\_freq\_file.tsv} available in the documentation.}
\label{tab:motif_prob}
\end{table}

\clearpage

\section{Extended results}

\subsection{Use case 1: \textit{de novo} motif discovery}
\label{Suppl:Use_Case_1}

The following parameters of inMOTIFin were used in this experiment.
The motif lengths were set between 4 and 9, with an average information content (IC) per position ranging from 0.1 to 1.8.
Ten motifs were created for each setting.
We simulated 1000 random background sequences of length 100 with equal and independent probabilities of sampling each base at each position.
In each simulation round, 50,000 sequences were sampled with replacement from the 1000 randomly created ones.
One motif instance was inserted in 95\% of these sequences.
The location of the motif instances was uniformly distributed across the sequence.
Each instance had forward orientation. 
Each simulation round was repeated 10 times with different random seeds.
On average, each motif was observed 4750 times.

To discover the injected motifs, RSAT was used with the following parameters.
We applied the following \textit{de novo} discovery algorithms: oligos, positions, and local\_words, with an oligo length of 6 to 7.
In each round, the number of motifs was set to 10, matching the known ground truth.
Thus, we evaluated the boundaries of discovery with slightly shorter and longer motifs than expected by the discovery algorithm.

In the comparison step, all motifs from the 10 rounds of each setting were pooled together.
This created a pool of 100 motifs.
For each round, the 10 discovered motifs were compared with the pool using Tomtom.
True positive motifs were those that were discovered and matched with the ground truth using Tomtom with an E-value threshold of 0.05.
False positives were motifs that were inserted but not found after motif comparison between the discovered and the ground truth.
False negatives were the motifs that matched some motif from the pool that was not inserted in the specific round.
True negatives were the motifs that were not found and indeed not inserted.

The results are shown in Supplementary Figure~\ref{fig:usecase1}.
Each circle is colored by the average Matthews Correlation Coefficient calculated across simulation rounds.
We can observe that longer motifs are easier to discover, even beyond the expected maximum length. However, even shorter nucleotide motifs as a minimum expectation are not found.
This is expected as the discovery algorithm sets a seed for each motif that can grow but not shrink.
Similarly, the higher average IC per position increased the probability of discovering the motif.
Interestingly, as low as 0.5 average IC per position was sufficient to find motifs when their length was at least 8.

\begin{figure}[tbh]
    \centering
    \includegraphics[width=1\linewidth]{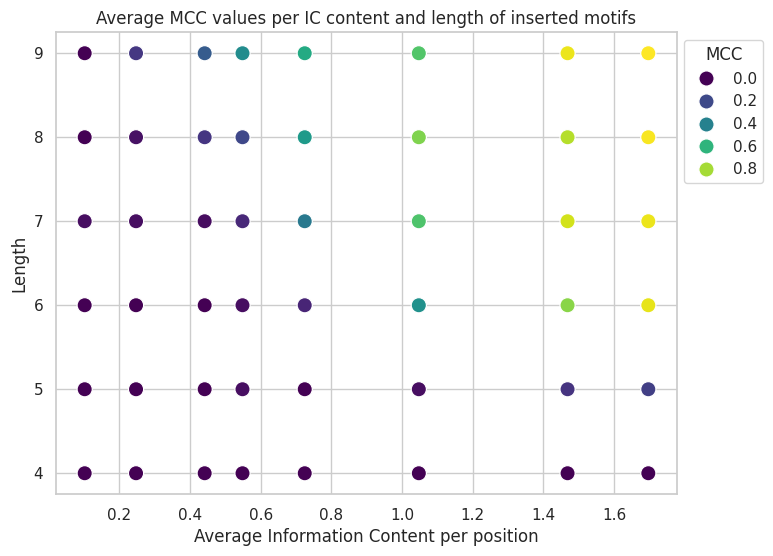}
    \caption{\textbf{Effect of IC content and length on the {\em de novo} motif discovery with RSAT.} }
    \label{fig:usecase1}
\end{figure}

\subsection{Use case 2: generating sequences with co-occurring motifs reflecting TF cooperativity}
\label{Suppl:Use_Case_2}

Several tools are available for the discovery of co-occurring or dimeric TF binding motifs. For instance, SpaMo identifies potential cooperative TFs within genomic regions~\cite{whitington_inferring_2011}. To perform this analysis, SpaMo requires a set of input sequences, each containing a primary motif positioned approximately at the center of the sequences. The software then searches upstream and downstream of the anchor motif to identify secondary motifs that are significantly enriched.

\begin{figure}[tbh]
    \centering
    \includegraphics[width=1\linewidth]{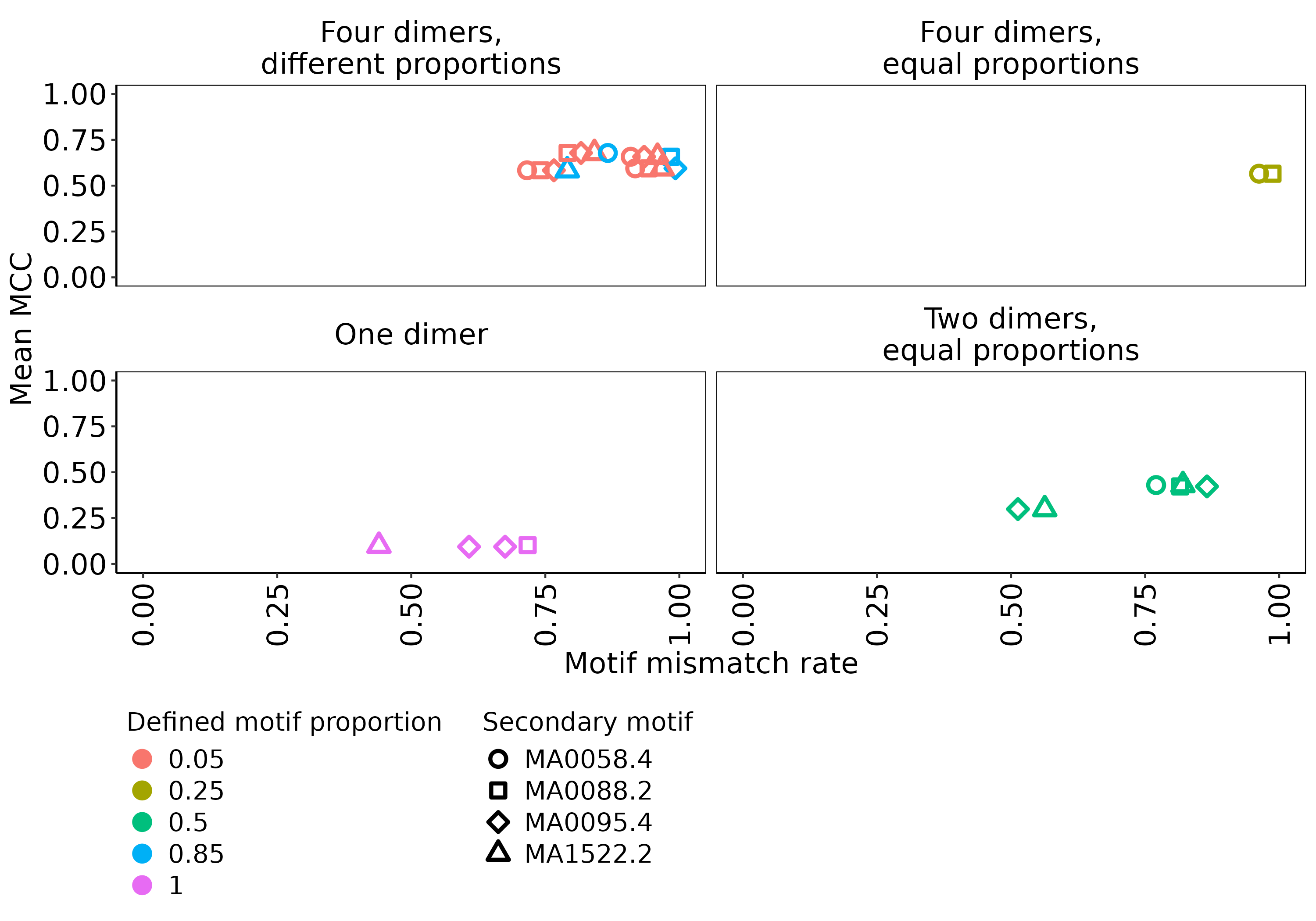}
    \caption{\textbf{Searching for secondary TF motifs from simulated sequences with inserted primary-secondary motif dimers}. The motif identity does not significantly influence the enrichment analysis results, whereas the presence of different dimers does. The example tool used in this analysis consistently identifies the target motifs as enriched, but recovers their instances more effectively when four dimers are inserted and less effectively when only one dimer is present.}
    \label{fig:usecase2}
\end{figure}

To simulate the input for this tool, we used inMOTIFin to create four dimeric motifs, where one part was identical in all (referred to as a primary motif). We chose the CTCF motif from the JASPAR database \cite{rauluseviciute_jaspar_2024} (motif ID MA0139.2).
Four other motifs, MAX (MA0058.4), YY1 (MA0095.4), MAZ (MA1522.2), and ZNF143 (MA0088.2), were chosen to represent the second motif of the dimer motifs.
We simulated random sequences with four dimers inserted with different proportions in the same number of sequences (6000 sequences, each case simulated ten times). The proportions ranged from 0 to 1, where some dimers were absent or inserted in all sequences.
The goal was to observe whether SpaMo could identify the secondary motif in the sites where it was inserted and how varying the proportions of the secondary motifs affects their discovery.
We repeated each insertion case 10 times and summarized using the Matthews correlation coefficient (MCC) (Supplementary Figure~\ref{fig:usecase2}).

The results in Supplementary Figure~\ref{fig:usecase2} show that all the motifs were significantly enriched in all cases. The identity of the secondary motif does not affect their recovery. However, there is a variation in MCC scores. The average MCC scores for 10 repeats were the highest when all four dimers were present in the set of sequences and lowest when only one dimer was inserted.
Notably, SpaMo has a high motif mismatch rate. The algorithm searches for enriched motifs using a collection of known motifs (in this case, we used JASPAR 2024 non-redundant CORE collection) \cite{rauluseviciute_jaspar_2024}. Therefore, multiple motifs representing binding of TFs from the same family will be found enriched.

Overall, with inMOTIFin sequence simulation, we can systematically evaluate motif enrichment tools by creating sequences with a variety of dimer motifs present.
Our analysis revealed that a tool can exhibit different performance depending on the motifs present and their frequency. This evaluation would be difficult, if not impossible, with real data.

\subsection{Use case 3: tool for explainability of deep learning models}
\label{Suppl:Use_Case_3}

We demonstrate the application of inMOTIFin for input perturbation analysis of a simple deep-learning (DL) model.
This model accepts 200 bp sequences as input and identifies the presence of GATA motifs, TAL motifs, or both through a final output layer consisting of three neurons.
We compare the model outputs between original background sequences and identical sequences with motifs inserted. Specifically, 1000 background sequences (200 bp each) were generated by shuffling real genomic sequences.
Motif insertion was then performed by inMOTIFin, producing three distinct sequence sets: sequences containing only GATA motifs, sequences containing only TAL motifs, and sequences containing both GATA and TAL motifs. TF binding motifs were directly obtained from the JASPAR 2024 non-redundant CORE collection (motif IDs MA0037.3 and MA0091.2, respectively).

\begin{figure}[tbh]
    \centering
    \includegraphics[width=1\linewidth]{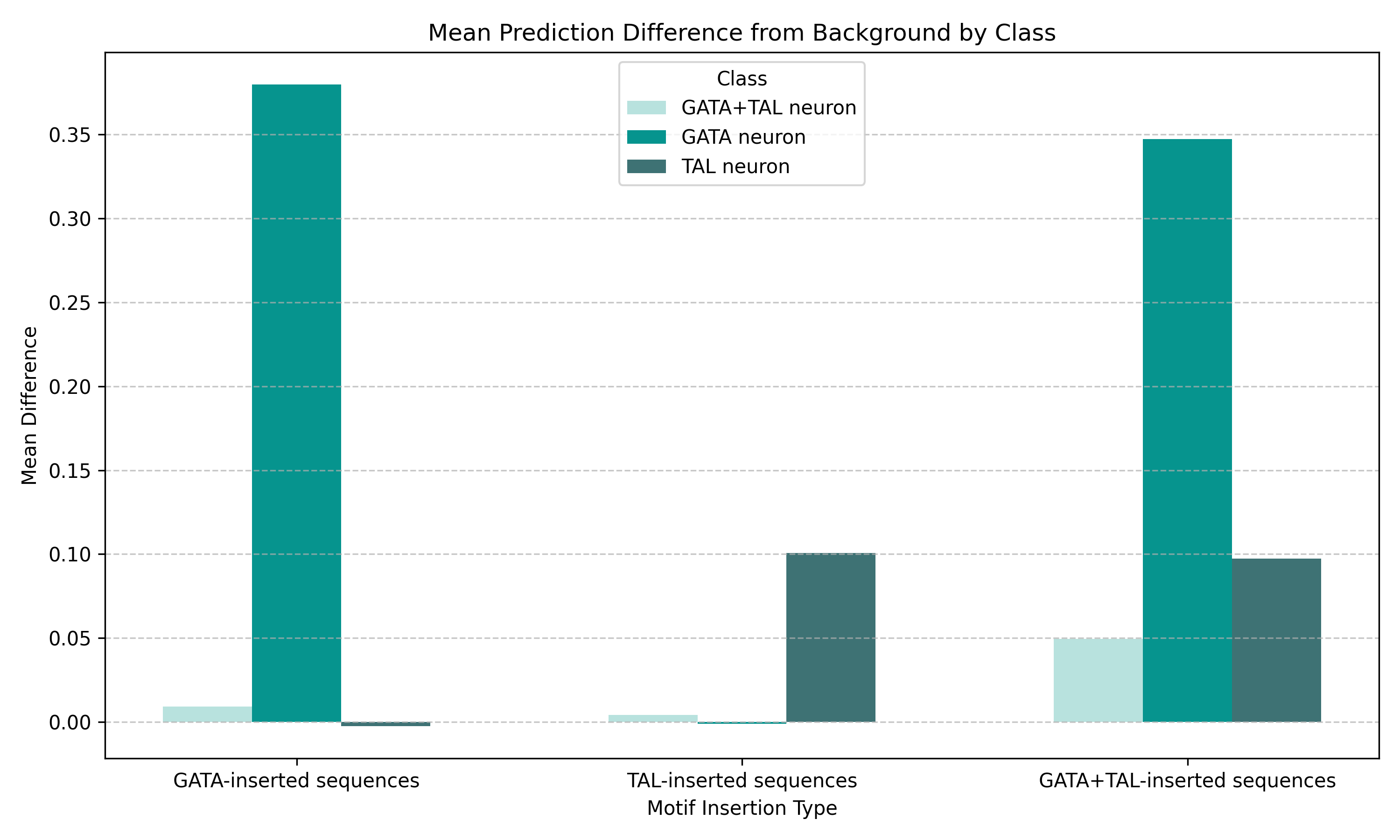}
    \caption{\textbf{Input perturbation of DL model, by motif insertion.} The insertion of GATA and/or TAL instances affects the output neurons with respect to the same background sequences.}
    \label{fig:usecase3}
\end{figure}

For sequences containing a single motif type, sampled motif instances were inserted precisely at the center of each sequence. For sequences containing both motifs, the two instances were inserted at uniformly distributed positions.
Consequently, we generated four sets of 1000 sequences each—one background and three motif-inserted sets—which were subsequently classified by the model. For each sequence, the model output was recorded as activation levels of the three output neurons.
We analyzed the differences in activation levels between each background sequence and its corresponding motif-inserted versions.
Figure~\ref{fig:usecase3} plots the average difference in activation across the 1000 sequences for each motif-inserted set, clearly illustrating how motif insertion alters the activation of the respective output neuron.

Variations in activation among the different motif-inserted sets reflect characteristics of the original model’s architecture and training conditions.
Notably, the TAL motif used in this analysis may differ slightly from the motif version used initially during model training, which accounts for the lower activation increase in the TAL-specific neuron.
Such analyses are valuable for understanding how the presence or absence of specific motifs impacts model predictions, thereby offering insights into underlying biological mechanisms.

\begin{figure}[tbh]
    \centering
    \includegraphics[width=\linewidth]{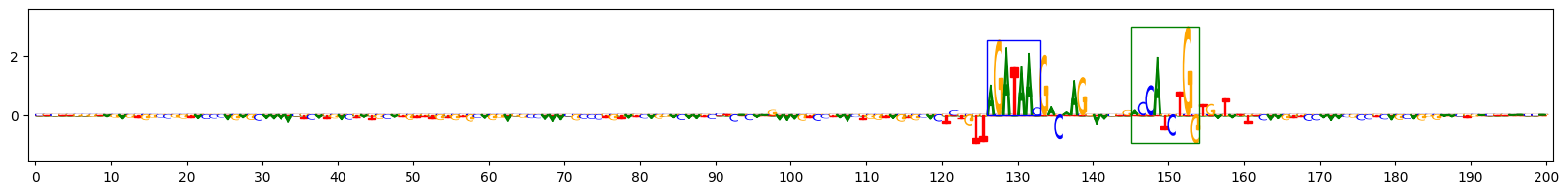}
    \includegraphics[width=\linewidth]{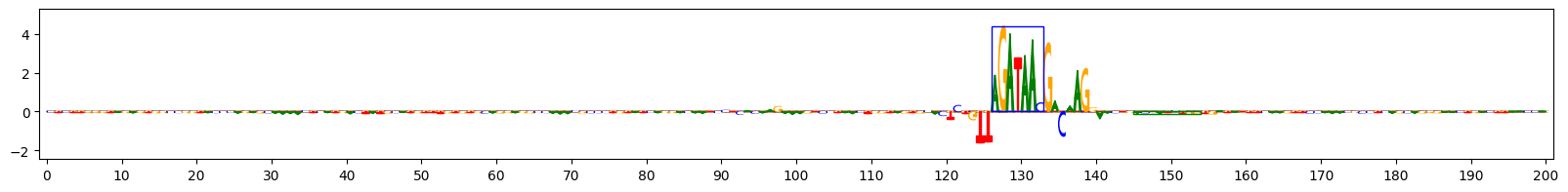}
    \includegraphics[width=\linewidth]{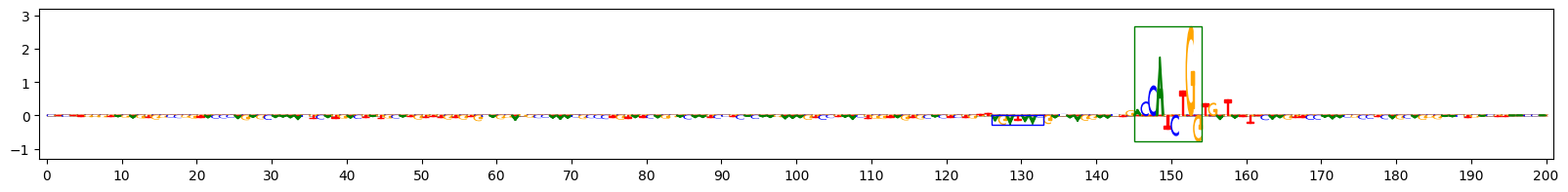}
    
    \caption{\textbf{Overview of DeepLift contribution scores}. The scores represent the results of the calculation performed on a sequence containing both GATA (blue box) and TAL (green box) motif instances. The calculation is performed on the GATA+TAL, GATA, and TAL neurons, respectively.}
    \label{fig:usecase3_deeplift}
\end{figure}

Other approaches to DL explainability include gradient-based attribution methods (e.g., SHAP \cite{lundberg2017unified} or DeepLift \cite{shrikumar_learning_2019}), which map the contributions to the output for each input feature.
For an input sequence, this means mapping them to each nucleotide. Analogously to the previous example, inMOTIFin assists in the creation of \textit{ad-hoc} sequences with inserted motifs to assess the contribution of nucleotides in motif instances to the model output.
This type of analysis helps highlight the base-specific importance score and is naturally comparable with motifs' PWM. Figure~\ref{fig:usecase3_deeplift} plots the scores obtained on a sequence with both motifs inserted, with respect to the three output neurons. The blue box highlights the GATA instance, and the green box highlights the TAL one. The coordinates are easily obtained from the inMOTIFin-generated BED file. The plots illustrate how the three neurons selectively recognize the bases associated with the two motifs. The TAL instance does not entirely match what it was trained on, as previously discussed. Again, inMOTIFin is a valuable tool for creating \textit{ad hoc} simulated sequences, which helps in understanding the model's behavior.

\clearpage

\section{Performance}

\begin{figure}[b]
    \centering
    \includegraphics[width=1\linewidth]{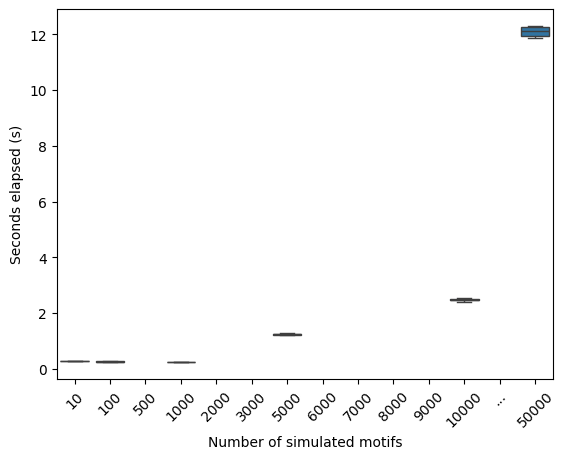}
    \caption{\textbf{Performance of motif simulation}. An increasing number of motifs of length within the range of 5 to 9 have been simulated.}
    \label{fig:performance1}
\end{figure}

\begin{figure}[t]
    \centering
    \includegraphics[width=1\linewidth]{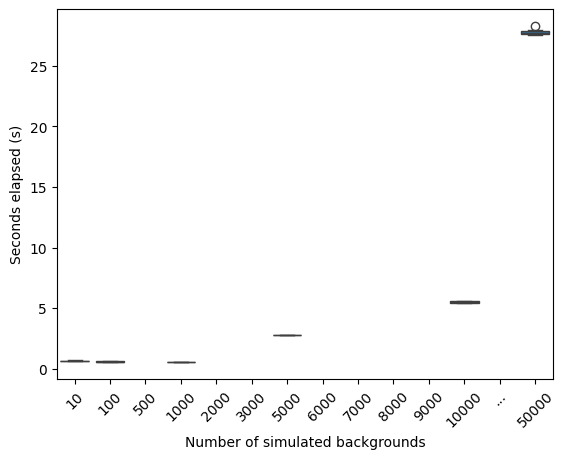}
    \caption{\textbf{Performance of background simulation.} An increasing number of background sequences of length 25 have been simulated.}
    \label{fig:performance2}
\end{figure}

\begin{figure}[b]
    \centering
    \includegraphics[width=1\linewidth]{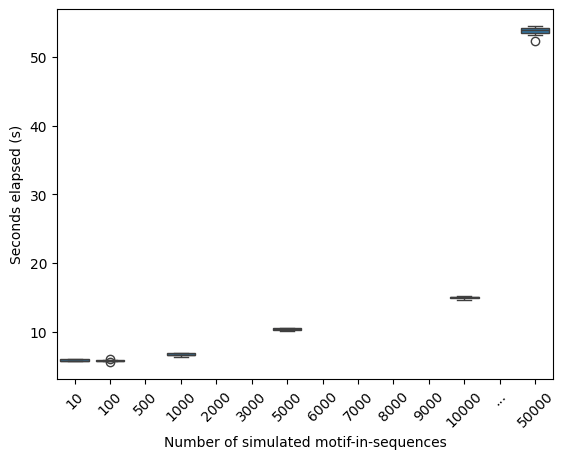}
    \caption{\textbf{Performance of motif-in-sequence simulation}. In the preparation phase, 50 motifs of length 5-9 were simulated and assigned to 5 groups of max size 10 (average size 7). The probability of groups being selected was assigned randomly, allowing for differences up to 8-fold. Within the groups, the probability of selection was equal for all motifs. Furthermore, 1000 backgrounds of length 250 were simulated. In the sampling phase, three groups and five motifs were selected per sequence. The positions of inserted motif instances were sampled from a Gaussian mixture model consisting of two distributions, centered at 50 and 210 bp, with variances of 10 and 3, respectively. The probability of motif instance orientation was 60\% forward and 40\% reverse.}
    \label{fig:performance3}
\end{figure}

The evaluation was performed on a desktop machine with an Intel® Core™ i7-10750H × 12 processor.
Supplementary Figures~\ref{fig:performance1}-\ref{fig:performance3} illustrate the outcome of this test.
The time complexity of each type of simulation is O(N).

\clearpage

\bibliographystyle{unsrt}
\bibliography{supplement}